\documentclass[conference]{IEEEtran}
\IEEEoverridecommandlockouts
\usepackage{cite}
\usepackage{empheq}
\usepackage{amsthm}
\usepackage{amsmath,amssymb,amsfonts,eqnarray}
\usepackage{graphicx}
\usepackage{textcomp}
\usepackage{xcolor}
\usepackage{acronym}
\usepackage{tabularx}
\usepackage{algorithm}
\usepackage{algpseudocode}
\usepackage{multirow}
\usepackage{academicons}
\usepackage{scalerel}
\usepackage{tikz}
\usetikzlibrary{svg.path}
\usepackage{subcaption}
\usepackage{url}
\usepackage{ragged2e}
\usepackage{steinmetz}
\usepackage[symbol]{footmisc}
\usepackage{mathtools}
\usepackage{bm}
\usepackage{makecell}
\setcellgapes{4pt}
\usepackage{enumitem}
\usepackage{siunitx}
\usepackage{glossaries}
\usepackage[font=scriptsize]{subcaption}
\usepackage[font=footnotesize]{caption}

\newacronym{ntn}{NTN}{non-terrestrial network}
\newacronym{ns3}{ns-3}{network simulator 3}
\newacronym{los}{LOS}{line-of-sight}
\newacronym{itu}{ITU}{International Telecommunication Union}
\usepackage{comment}

\newcommand{\norm}[1]{\left\lVert#1\right\rVert}
\def\BibTeX{{\rm B\kern-.05em{\sc i\kern-.025em b}\kern-.08em
    T\kern-.1667em\lower.7ex\hbox{E}\kern-.125emX}}

  \tikzset{
  orcidlogo/.pic={
    \fill[orcidlogocol] svg{M256,128c0,70.7-57.3,128-128,128C57.3,256,0,198.7,0,128C0,57.3,57.3,0,128,0C198.7,0,256,57.3,256,128z};
    \fill[white] svg{M86.3,186.2H70.9V79.1h15.4v48.4V186.2z}
                 svg{M108.9,79.1h41.6c39.6,0,57,28.3,57,53.6c0,27.5-21.5,53.6-56.8,53.6h-41.8V79.1z M124.3,172.4h24.5c34.9,0,42.9-26.5,42.9-39.7c0-21.5-13.7-39.7-43.7-39.7h-23.7V172.4z}
                 svg{M88.7,56.8c0,5.5-4.5,10.1-10.1,10.1c-5.6,0-10.1-4.6-10.1-10.1c0-5.6,4.5-10.1,10.1-10.1C84.2,46.7,88.7,51.3,88.7,56.8z};
  }
}

\newcommand\orcidicon[1]{\href{https://orcid.org/#1}{\mbox{\scalerel*{
\begin{tikzpicture}[yscale=-1,transform shape]
\pic{orcidlogo};
\end{tikzpicture}
}{|}}}}
\usepackage{hyperref}

\def\BibTeX{{\rm B\kern-.05em{\sc i\kern-.025em b}\kern-.08em
    T\kern-.1667em\lower.7ex\hbox{E}\kern-.125emX}}
\begin{document}
\definecolor{orcidlogocol}{HTML}{A6CE39}

\title{Preliminary Performance Evaluation of a Satellite-to-HAP Communication Link}

     
\author{\IEEEauthorblockN{ Giovanni Grieco$^{\circ }$, Giovanni Iacovelli$^{\circ }$, Mattia Sandri$^{\star }$, Marco Giordani$^{\star }$, Michele Zorzi$^{\star }$, Luigi Alfredo Grieco$^{\circ }$\medskip}
\IEEEauthorblockA{
$^{\circ}$Department of Electrical and Information Engineering, Politecnico di Bari, Italy\\
Email: \{giovanni.grieco;giovanni.iacovelli;alfredo.grieco\}@poliba.it\vspace{0.1cm} \\
$^{\star}$Department of Information Engineering, University of Padova, Italy \\ (
Email: \{sandrimatt;giordani;zorzi\}@dei.unipd.it)}}

\acrodef{5G}{Fifth-Generation}

\acrodef{GEO}{Geostationary Equatorial Orbit}

\acrodef{HAP}{High Altitude Platform}

\acrodef{IoD}{Internet of Drones}
\acrodef{IoD-Sim}{Internet of Drones Simulator}
\acrodef{IoT}{Internet of Things}
\acrodef{ITU}{International Telecommunication Union}

\acrodef{JSON}{JavaScript Object Notation}

\acrodef{ns-3}{Network Simulator 3}
\acrodef{NTN}{Non-Terrestrial Network}

\acrodef{PoI}{Point of Interest}
\acrodefplural{PoI}{Points of Interest}

\acrodef{SNR}{Signal-to-Noise Ratio}

\acrodef{UAV}{Unmanned Aerial Vehicle}

\acrodef{VLC}{Visible Light Communications}

\newcommand{\TODO}{\textcolor{blue}{TODO}}
\newcounter{mytempeqncnt}

\renewcommand{\qedsymbol}{\scalebox{0.75}{$\blacksquare$}}
\newtheorem{assumption}{Assumption}
\newtheorem{remark}{Remark}
\newtheorem{theorem}{Theorem}
\newtheorem{corollary}{Corollary}
\newtheorem{lemma}{Lemma}

\maketitle

\begin{abstract}
The emergence of \ac{5G} communication networks has brought forth unprecedented connectivity with ultra-low latency, high data rates, and pervasive coverage. 
However, meeting the increasing demands of applications for seamless and high-quality communication, especially in rural areas, requires exploring innovative solutions that expand \ac{5G} beyond traditional terrestrial networks. Within the context of \acp{NTN}, two promising technologies with vast potential are \acp{HAP} and satellites. The combination of these two platforms is able to provide wide coverage and reliable communication in remote and inaccessible areas, and/or where terrestrial infrastructure is unavailable.
This study evaluates the performance of the communication link between a \ac{GEO} satellite and a \ac{HAP} using the \ac{IoD-Sim}, implemented in ns-3 and incorporating the 3GPP TR 38.811 channel model. 
The code base of \ac{IoD-Sim} is extended to simulate \acp{HAP}, accounting for the Earth's curvature in various geographic coordinate systems, and considering realistic mobility patterns.
A simulation campaign is conducted to evaluate the GEO-to-HAP communication link in terms of \ac{SNR} in two different scenarios, considering the mobility of the HAP, and as a function of the frequency and the distance.
\end{abstract}


\acresetall

\section{Introduction}\label{sec:introduction}
Fifth-generation (5G) wireless networks \cite{7414384} started a new era in terms of connectivity by promising ultra-low latency, high data rates, and ubiquitous coverage.
Still, as the demands for seamless, pervasive, and high-quality communications grow, innovative solutions are being explored to extend 5G beyond traditional terrestrial networks.
Notably, \acp{NTN}~\cite{9275613} are capable of bridging geographical divides, and provide broadband standalone connectivity even in the absence of terrestrial infrastructures (e.g., in rural or remote areas) or when terrestrial infrastructures are unavailable (e.g., in case of emergency). In the context of \ac{NTN}, two key technologies that hold very high potential are \acp{HAP} and satellites. 

\acp{HAP}, also known as stratospheric platforms, are \acp{UAV} soaring in the stratosphere at altitudes ranging from $20$ to $\SI{50}{km}$. These platforms can be equipped with propulsion systems, typically based on propellers and electric motors, to move to different locations~\cite{GARCIAGUTIERREZ2020105562}. This capability allows them to be deployed as needed, providing coverage to specific areas or addressing changing communication demands. Moreover, \acp{HAP} can establish wireless links with satellites, other \acp{HAP}, low-altitude \acp{UAV} such as drones, and/or terrestrial networks. Indeed, flying at high altitudes, \acp{HAP} can offer wide coverage and great line-of-sight connections, and establish reliable communication links in previously inaccessible regions.
In turn, satellites have been used for decades, primarily for navigation, meteorology, or television broadcasting. However, with the advent of 5G, satellites are now considered as an integral part of the communication infrastructure, to support cost-effective, high-capacity, wide-coverage connectivity on the ground~\cite{8930826,9999287}.

\begin{figure*}
    \centering
    \includegraphics[width=\textwidth]{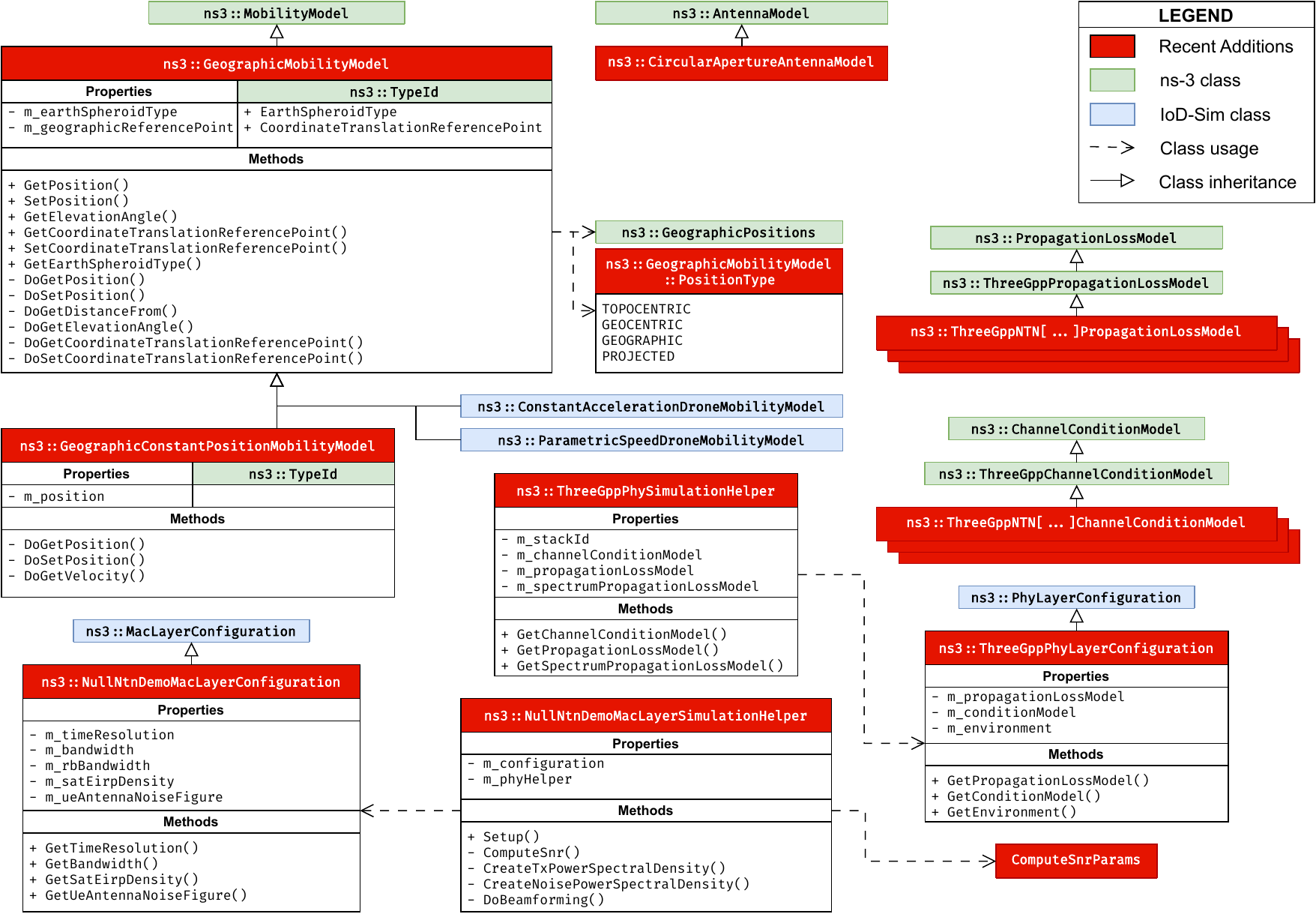}
    \caption{Class diagram of the recent additions introduced in IoD-Sim to simulate HAP-to-satellite communication.}
    \label{fig:class-diagram}
\end{figure*}

The integration of \acp{HAP} and satellites into the 5G ecosystem brings several advantages. First, these aerial and space platforms can effectively bridge the digital divide by bringing high-speed connectivity to remote areas where ground infrastructure is limited or absent~\cite{chaoub20216g}.
Moreover, \acp{HAP} and satellites can play a key role in disaster response and recovery scenarios, providing emergency communication networks when terrestrial infrastructure is (temporally or permanently) disrupted or unavailable.
Furthermore, 
as the demands of data-hungry applications increase, \acp{HAP} and satellites can supplement existing terrestrial networks, and relieve congestion by offloading traffic. Additionally, they can support critical applications requiring ubiquitous and uninterrupted connectivity, such as for autonomous vehicles, smart cities, and \ac{IoT} devices that operate in remote or mobile environments. However, we claim that this potential can be maximized if HAPs and satellites work together as a multi-layered integrated network~\cite{wang2021potential}, rather than as standalone solutions. For example, the HAP layer can act as a wireless relay to improve the link quality of an upstream satellite. At the same time, the satellite layer can offer the HAP a ready-to-use link for the backhaul, as well as an easy access to the core network. However, it is still unclear whether satellite-to-HAP communication is feasible and, if so, how it can be realized, which motivates our study.

In this context, the \ac{IoD-Sim}~\cite{iodsim} is a comprehensive simulation platform for the \ac{IoD} \cite{BSG21}, which extends the \ac{ns-3} code base with additional features to simulate \ac{IoD} networking elements (e.g., drones, network access points), entities, mobility models.
The scenario configuration can be easily defined in \ac{JSON} by the user, and does not require particularly advanced coding expertise.
Given its flexible and modular structure, it represents a valuable tool to design and evaluate \ac{NTN} scenarios. 


In light of the above, this work evaluates the PHY-layer performance of the communication link between a \ac{HAP} and a \ac{GEO} satellite.
With this aim, \ac{IoD-Sim} has been extended to incorporate: (i) the channel model, standardized in 3GPP TR 38.811~\cite{38811} and implemented in the \texttt{ns3-ntn} module~\cite{sandrintnns3}, to simulate HAP-to-satellite communication; (ii) HAP-specific mobility models that take into account the impact of the Earth's curvature; and (iii) new coordinate systems, i.e., geographic, geocentric, topocentric, and projected, to facilitate object placement and mobility.
To assess this preliminary implementation, \ac{IoD-Sim} and its extensions are tested in terms of \ac{SNR} as a function of the distance between the \ac{HAP} and the \ac{GEO} satellite and of the frequency.
We consider two different scenarios with a real satellite position and (i) a \ac{HAP} that moves from northern Europe to central Africa, and (ii) a \ac{HAP} that hovers below the satellite.

Notice that, even though the current version of \ac{IoD-Sim} primarily focuses on channel and physical layer aspects, it can be readily integrated with the rest of the ns-3 protocol stack, and therefore represents a crucial tool to support more advanced end-to-end protocol design and evaluations in the context of \ac{NTN}.

The remainder of the paper is organized as follows. 
Sec.~\ref{sec:sysmod} describes our system and channel models, Sec.~\ref{sec:evaluation} presents the simulation campaign and numerical results, and Sec.~\ref{sec:conclusions} concludes the paper with final suggestions for future work.

\section{System and Channel Model Implementation}\label{sec:sysmod}
In this section we describe our system (Sec.~\ref{sec:sysimpl}) and channel (Sec.~\ref{sec:channel}) models, and their relative implementation in \ac{IoD-Sim} according to the structure in Figure \ref{fig:class-diagram}. 

\subsection{System Model Implementation}\label{sec:sysimpl}
In this paper, GEO-to-HAP communication is referred to as a ``mission'' of $T$ seconds, discretized into $K$ time slots of equal duration $\delta$. As a consequence, the \ac{HAP} pursues a trajectory embodied by a set of discrete points, each expressed in terms of latitude, longitude, and altitude by vector $\textbf{q}_k \in \mathbb{R}^3$, with $k=1,\ldots,K$. Similarly, the \ac{GEO} satellite is located at $\textbf{w} \in \mathbb{R}^3$, which obviously does not change over time.
While latitude and longitude are expressed in radians, altitude is expressed in meters.

For the sake of practicality, the trajectory curve of the \ac{HAP} is generated leveraging a revised version of the original B\'ezier equation. 
It is defined by a set of $N$ \acp{PoI}, denoted as $\textbf{P} = \bigl\{ \textbf{p}_0, \textbf{p}_1, \dots, \textbf{p}_{N-1} \bigr\}$, with $\textbf{p}_i \in \mathbb{R}^3$ expressed in geographic coordinates.
These \acp{PoI} are then projected over a Cartesian space, known as the EPSG:3857 WGS84/Pseudo-Mercator projection \cite{EPSG:3857}, which is defined as
\begin{align}
    x &= \frac{2^{\alpha}}{2\pi} (\lambda + \pi), \\
    y &= \frac{2^{\beta}}{2\pi} \left(\pi - \ln\left(\tan\left(\frac{\pi}{4} + \frac{\varphi}{2}\right)\right)\right), \\
    z &= z,
\end{align}
where $\lambda$ is the latitude, $\varphi$ is the longitude,  $\alpha = 25.059$, and $\beta = 24.665$. The latter two constants are used to normalize the Cartesian space's unit of measurement to meters.

Each \ac{PoI} is assigned a level of interest in $\textbf{l} = \bigl\{ l_0, l_1, \dots, l_{N-1} \bigr\}$. 
 As the level of interest of a certain \ac{PoI} grows, the resulting trajectory of the HAP is configured to pass closer to the \ac{PoI} itself, and can be expressed~as 
\begin{equation}
    \mathbf{\overline{q}}_k = \sum_{i=0}^{N-1} \mathbf{\overline{p}}_i \sum_{j=0}^{l_i - 1} \binom{\Lambda}{L_i + j} (1-t)^{\Lambda-L_i-j} t^{L_i+j},
    \label{eq:trajectoryGenerator}
\end{equation}
with $t = k/K$, while $\mathbf{\overline{q}}_k, \forall k$, and $\mathbf{\overline{p}_i}, \forall i$, represent the Mercator projected coordinates \cite{EPSG:3857}. Moreover, $\Lambda = \left( \sum_{i=0}^{N-1} l_i \right) - 1$ and $L_i = \sum_{h=0}^{i-1} l_h$ are used as auxiliary variables.

Notably, multiple B\'ezier curves can be defined to force this behavior, which only requires the geographical coordinates of the \acp{PoI}. 
In \ac{IoD-Sim}, this mobility model is implemented in the \texttt{ns3::ParametricSpeedDroneMobilityModel} class, which was extended from the original code base in \cite{iodsim} with the boolean attribute \texttt{UseGeographicSystem}, and in the \texttt{ns3::ConstantAccelerationDrone\-MobilityModel} class.

To evaluate the distance between two nodes at different heights, another Cartesian system should be used, i.e., the geocentric one. This reference system has its point of origin at the center of the Earth. To this end, the geographic coordinates $\mathbf{q}_k, \forall k$, and $\mathbf{w}$ have been transformed using the WGS84 ellipsoid. First, we compute the polar radius as
\begin{align}
    r = \frac{a}{\sqrt{1 - e^2 \sin^2(\lambda)}},
\end{align}
where $a = 6378137$ and $e = 0.0818191908426215$ are the Earth's semi-major axis and its eccentricity, respectively. Then, the points in geocentric coordinates are expressed as:
\begin{align}
     x^\prime &= (r + z) \cos(\lambda) \cos(\varphi), \\
     y^\prime &= (r + z) \cos(\lambda) \sin(\varphi), \\
     z^\prime &= ((1 - e^2) r + z) \sin(\lambda).
\end{align}
Therefore, the \ac{HAP}-satellite distance, which accounts for the curvature of the Earth in each time slot, can be expressed as $d_k = \norm{\textbf{q}'_k-\textbf{w}'}$.

As far as the system model is concerned, PHY-layer parameters can be set in \ac{IoD-Sim} prior to the simulation via \texttt{ns3::ThreeGppPhySimulationHelper}, which is configured in the JSON file by \texttt{ns3::ThreeGppLayerConfiguration}, as illustrated in Figure \ref{fig:class-diagram} and as thoroughly described in \cite[Sec. V-F]{iodsim}.
The same logic is applied at the MAC layer via \texttt{ns3::NullNtnDemoMacLayerSimulationHelper}, where  \texttt{ns3::NullNtnDemoMacLayerConfiguration} is also designed to verify that the PHY layer acts according to the reference standard.

\subsection{Channel Model Implementation}
\label{sec:channel}
The \ac{HAP}-to-satellite communication link is modeled according to the 3GPP TR 38.811 specifications~\cite{38811}, which are in turn based on the cellular channel model presented in 3GPP TR 38.901~\cite{3gpp.38.901}. A first characterization of the above has also been implemented in \ac{ns-3} in the \texttt{ns3-ntn} module~\cite{sandrintnns3}, and eventually extended into the current version of \ac{IoD-Sim} in the \texttt{ns3::ThreeGppNTN[...]ChannelConditionModel} and \texttt{ns3::ThreeGppNTN[...]Propagation\-LossModel} classes, as represented in Figure \ref{fig:class-diagram}.
Specifically, the simulator supports different 3GPP channel environments, i.e., dense urban, urban, suburban, and rural.

The channel model accounts for several attenuation factors: basic path loss, atmospheric absorption, and scintillation.

\paragraph{Basic path loss} 
It is characterized by three main components, and can be written as
\begin{equation}
    PL_{b} = FSPL+SF+CL,
\label{eq:pl}
\end{equation}
where $FSPL$ is the free-space path loss, $SF$ is the shadow fading, and $CL$ represents the clutter loss. The free-space path loss for the \ac{NTN}
scenario is defined as
\begin{equation}
    FSPL = 32.45 + 20\log_{10}(f_{c}) + 20\log_{10}(d_k),
    \label{eq:fspl}
\end{equation}
where $f_{c}$ is the carrier frequency in GHz, and $d_k$ is the distance between the HAP at generic point $k$ in its trajectory, and the satellite in meters. 
The shadow fading is modeled as a log-normal random variable, i.e.,
\begin{equation}
  SF \sim N\left ( 0,\sigma_{SF}^{2} \right ), 
\end{equation}
and depends on the elevation angle, the visibility condition (i.e., line-of-sight or not), and the carrier frequency. 
The characterization of the clutter loss follows a similar model, even though it can be usually neglected in line-of-sight.


\paragraph{Atmospheric absorption} 
Unlike in the terrestrial channel, in the NTN scenario the propagation of the signal undergoes an additional attenuation due to penetration through the atmosphere. 
While the complete model is available in the  \ac{ITU} documents~\cite{itup676}, the 3GPP TR 38.811 specifications adopt a simplified model considering only the annual mean values of absolute humidity, water-vapor density, water-vapor partial pressure, and dry air pressure for the atmosphere. Therefore, atmospheric absorption is given~by
\begin{equation}
    PL_{A}=\frac{A_{\rm zenith} }{\sin(\xi)},
\end{equation}
where $A_{\rm zenith}$ is the absorption loss in dB at the zenith angle at a given carrier frequency, and $\xi$ is the actual elevation angle. The value of $A_{\rm zenith}$ is given in \cite{itup676}. 
Atmospheric absorption is relevant only for frequencies above $\SI{10}{GHz}$, or in the case of an elevation angle lower than 10 degrees for all frequencies. An important absorption effect is due to the presence of oxygen, which produces a very significant attenuation at frequencies around $\SI{60}{GHz}$ \cite[Annex 1.1]{itup676}.

\begin{table}
\centering
\renewcommand{\arraystretch}{1.4}
\caption{Simulation parameters and settings.}
\label{tab:simparams}
\centering
\begin{tabular}{|l|l|}
\hline
{Parameter} & {Value} \\ \hline
Mission duration ($T$) & $\SI{284}{[h]}$ \\ 
3GPP Environment & \ac{NTN} Rural~\cite[Sec.6.1.2]{38811} \\ 
Update period & $\SI{1}{[s]}$ \\ 
Frequency ($f_c$) & $\SI{20}{[GHz]}$ \\ 
Shadowing & Disabled \\ 
Time resolution & $\SI{1000}{[s^{-1}]}$ \\ 
Bandwidth & $\SI{400}{[MHz]}$ \\ 
EIRP density & $\SI{40}{[dBW/MHz]}$ \\ 
Antenna noise figure & $\SI{1.2}{[dB]}$ \\ 
\ac{HAP} speed & $\SI{24}{[m/s]}$  \\ 
GEO antenna gain & $\SI{58.5}{[dBi]}$ \\ 
\ac{HAP} antenna Gain & $\SI{39.7}{[dBi]}$ \\ 
GEO antenna radius & $\SI{2.5}{[m]}$ \\ 
\ac{HAP} antenna radius & $\SI{0.3}{[m]}$ \\ 
GEO antenna inclination & $\SI{180.0}{[deg]}$ \\ 
\ac{HAP} antenna inclination & $\SI{0}{[deg]}$ \\ 
1st \ac{PoI} (Takeoff/Landing) & $[78.244789^\circ,\,15.4843571^\circ,\,\SI{20}{km}]$ \\ 
2nd \ac{PoI} (Iran \ac{PoI}) & $[35.7074505^\circ,\,51.1498211^\circ,\,\SI{20}{km}]$ \\ 
3rd \ac{PoI} (\ac{GEO} Satellite) & $[0.04,\,-4.95,\,\SI{20}{km}]$ \\ 
4th \ac{PoI} (Iceland \ac{PoI}) & $[64.133542^\circ,\,-21.9348416^\circ,\,\SI{20}{km}]$ \\ \hline
\end{tabular}
\end{table}

\begin{figure}
    \centering
    \includegraphics[width=\columnwidth]{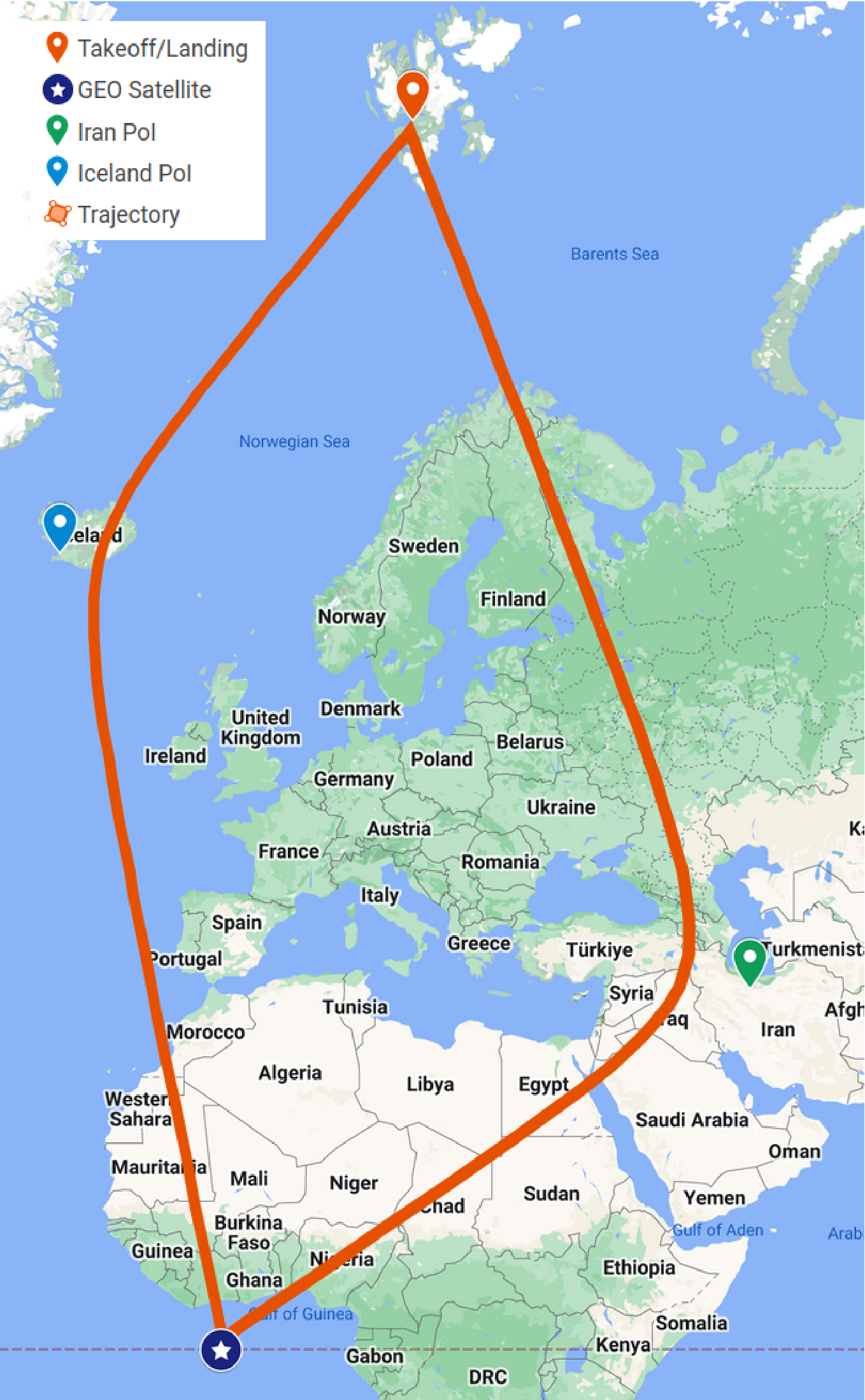}
    \caption{An overview of the trajectory of the \ac{HAP}, its \acp{PoI}, and the satellite position over the Earth.}
    \label{fig:traj-big-withlegend}
\end{figure}

\paragraph{Scintillation} 
It determines the rapid fluctuations of the phase and the amplitude of the signal, caused by small-scale changes in the structure of the atmosphere.
Specifically, scintillation is due to two different contributions: 
{tropospheric scintillation} and {ionospheric scintillation}. 
The former is particularly significant for frequencies above 10 GHz and at low elevation due to the longer path of the signal. It is modeled as the 99-percentile of the attenuation level observed in Toulouse (France) at $\SI{20}{GHz}$, as reported in~\cite[Figure~6.6.6.2.1-1]{38811}.
Ionospheric scintillation, instead, is relevant only for latitudes below 20 degrees, or for frequencies below 6 GHz. It is expressed as
\begin{equation}
    PL_{IS} = \left ( \frac{f_{c}}{4} \right )^{-1.5}\frac{P_{\rm fluc}\left ( 4 \text{ GHz} \right )}{\sqrt{2}},
\end{equation}
where $P_{\rm fluc}\left ( 4 \text{ GHz} \right )$ represents the ionospheric attenuation level 
observed in Hong Kong between March 1977 and March 1978 at a frequency of 4 GHz~\cite[Figure~6.6.6.1.4-1]{38811}. 


\section{Simulation Campaign}
\label{sec:evaluation}

\begin{figure}[ht]
    \centering
    \includegraphics[width=\columnwidth]{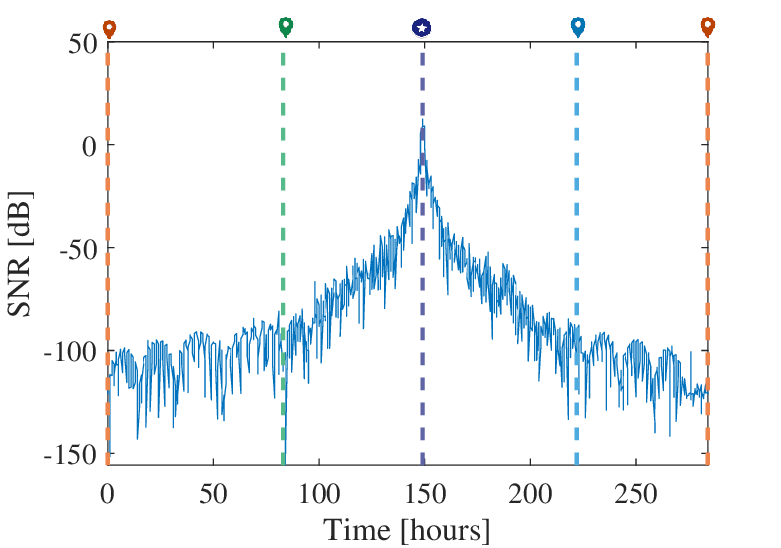}
    \caption{The evolution of the \ac{SNR} during the mission.}
    \label{fig:traj-big-snr}
\end{figure}

In this section, we evaluate via simulation the channel link between a GEO satellite and a \ac{HAP}. The satellite is located at $[0.04^\circ,\,-4.95^\circ, \SI{35770.88}{km}]$, which corresponds to the actual position of Eutelsat 5 West B. The \ac{HAP} follows a curvilinear trajectory generated with 4 \acp{PoI}. Besides, the \ac{HAP} adopts the mobility model described in Sec.~\ref{sec:sysimpl}, with a constant speed of $\SI{24}{m/s}$. This leads to a total mission duration of $T = \SI{1023160}{s}\simeq284$ h. A comprehensive overview of the described scenario and mobility pattern is illustrated in Figure~\ref{fig:traj-big-withlegend}, while simulation parameters are listed in Table~\ref{tab:simparams}.

The HAP and the GEO satellite are equipped with a circular aperture antenna operating at $\SI{20}{GHz}$. This antenna, also known as reflector, is modeled based on the \texttt{ns3::CircularApertureAntennaModel} class in the \texttt{ns3-ntn} module~\cite{sandrintnns3}. The \ac{HAP} (GEO satellite) antenna has a maximum gain of $\SI{39.7}{dB}$ ($\SI{58.5}{dB}$), a diameter of $\SI{0.6}{m}$ ($\SI{5}{m}$), and an inclination angle of $0^\circ$ ($180^\circ$). We focus on downlink communication, where signals are sent from the satellite to the \ac{HAP} with a transmission power of $\SI{37.5}{dBm}$ and a bandwidth of $\SI{400}{MHz}$.

We consider the channel model described in Sec.~\ref{sec:channel}, and a rural environment~\cite[Sec.6.1.2]{38811} with the assumption of line-of-sight visibility.
Given that the \ac{HAP} flies in the stratosphere, i.e., at a fixed altitude of $\SI{20}{km}$, we assume that the impact of shadowing as well as of tropospheric scintillation is negligible. Moreover, we consider the impact of atmospheric absorption through all the layers of the atmosphere, even though the \ac{HAP} flies in the stratosphere, so as to obtain worst-case results.

Considering the above setup, in Figure \ref{fig:traj-big-snr} we illustrate the evolution of the \ac{SNR} over time of the link between the GEO satellite and the HAP, during the mission. The markers refer to those in Figure \ref{fig:traj-big-withlegend}, and indicate when the \ac{HAP} reaches a certain \ac{PoI} according to the given level of interest.
As expected, as the \ac{HAP} approaches the geographical position of the GEO satellite (i.e., the starred marker, corresponding to the area in the Gulf of Guinea), the \ac{SNR} increases, thus reaching a maximum value of $\SI{13.0584}{dB}$. 
With a bandwidth of 400 MHz as per the 3GPP TR 38.811 specifications, this corresponds to a PHY-layer capacity of approximately $\SI{1.78}{Gbps}$, which is enough to realize HAP-to-satellite communication. 
However, the \ac{SNR} drops below $\SI{0}{dB}$ as the HAP moves farther away from the GEO satellite, i.e., as the length of the link between the two endpoints increases.

\begin{figure}
    \centering
    \includegraphics[width=\columnwidth]{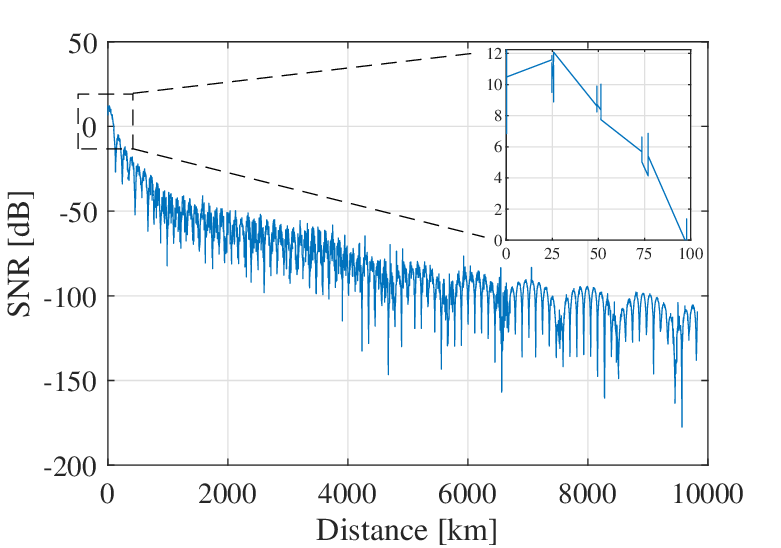}
    \caption{\ac{SNR} vs. the distance between the \ac{HAP} and the \ac{GEO} satellite, projected on the Earth.}
    \label{fig:dist-snr}
\end{figure}

\begin{figure}
    \centering
    \includegraphics[width=\columnwidth]{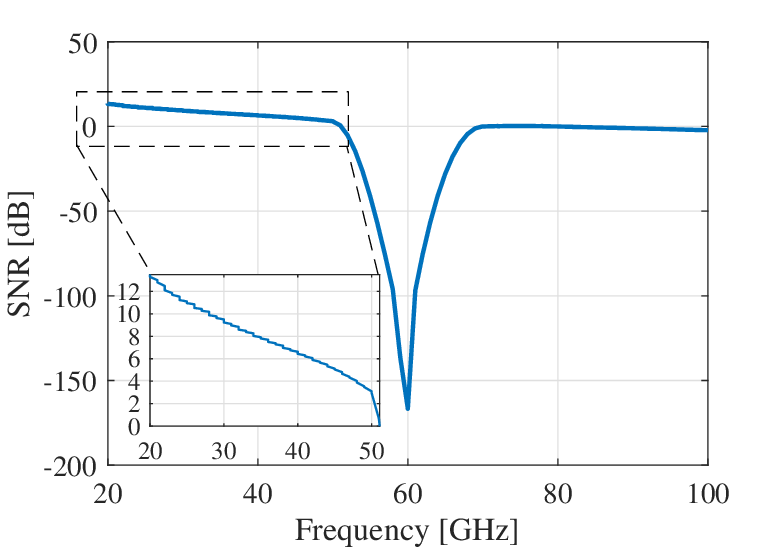}
    \caption{\ac{SNR} in the PoI of maximum link gain vs. the frequency.}
    \label{fig:freq-snr}
\end{figure}

For additional insights, Figure \ref{fig:dist-snr} shows the SNR as a function of the distance between the HAP and the GEO satellite projected over the Earth. As expected, the \ac{SNR} is positive only for distances lower than $\SI{\sim100}{km}$, which roughly corresponds to the service area of the \ac{HAP}, and then drops below $\SI{0}{dB}$ everywhere else. This is due to (i) the high directivity of reflector antennas, which poses a limit to the coverage radius of the HAP, and (ii) the higher path loss as the distance between the HAP and the GEO satellite increases, and the elevation angle between the two decreases accordingly.

Finally, we analyze a scenario in which the \ac{HAP} hovers below the \ac{GEO} satellite in the PoI of maximum link gain (i.e., the starred marker in Figure~\ref{fig:traj-big-withlegend}), and the frequency varies from $20$ to $\SI{100}{GHz}$. We can see in Figure~\ref{fig:freq-snr} that the \ac{SNR} decreases as the frequency increases, as the $FSPL$ in Eq.~\eqref{eq:fspl} increases, with a significant drop around $\SI{60}{GHz}$ due to the impact of oxygen absorption in the atmosphere (in the order of $\SI{15}{dB/km}$). Still, the SNR is consistently above $\SI{0}{dB}$ as ${f_c\leq 50}$ GHz, where the very large bandwidth at these frequencies can support high-rate transmissions.

In conclusion, the above results demonstrate that \ac{NTN} communication between a GEO satellite and a HAP can be effectively established, at least from a PHY-layer standpoint, and simulated using \ac{IoD-Sim}.

\section{Conclusions and Future Work}\label{sec:conclusions}
This work presents a preliminary implementation in \ac{IoD-Sim} of the channel model between a HAP and a GEO satellite, as per the 3GPP TR 38.811 specifications. To do so, \ac{IoD-Sim} has been extended to simulate new mobility models for the HAP, and now also supports geocentric, geographic, topocentric, and projected coordinates. 
Simulation results show that high-capacity GEO-HAP communication is feasible, and we provide indications on the optimal set of frequencies and distances for maximum performance.

Several aspects are still not yet addressed in the simulator. Specifically, future implementation efforts will be focused on the modeling of (i) non-stationary satellite orbits, (ii) \ac{HAP} and satellite power consumption, (iii) MAC-layer protocols that take into account \ac{NTN} propagation delays, and (iv) an Integrated-T/NTN end-to-end communication stack for a comprehensive 6G simulation platform.

\section*{Acknowledgments}
This work was partially supported by the European Union under the Italian National Recovery and Resilience Plan (NRRP) of NextGenerationEU, with particular reference to the partnership on ``Telecommunications of the Future'' (PE00000001 - program ``RESTART''). 

\bibliographystyle{IEEEtran}
\bibliography{bibliography}

\end{document}